\documentclass[aps,prl,showpacs,showkeys,
twocolumn,superscriptaddress,altaffilletter,
 amsmath,amssymb]{revtex4-1}
\usepackage{graphicx}
\usepackage{dcolumn}
\usepackage{bm}
\usepackage[mathlines]{lineno}
\usepackage{cleveref}
\usepackage{comment}
\usepackage{amssymb}     
\usepackage{marvosym}    
\usepackage{upgreek}     
\newcommand{\kgy}         {{kg$\cdot$yr}}
\newcommand{\kgyr}        {{kg$\cdot$yr}}

\newcommand{\Qbb}         {{$Q_{\beta\beta}$}}

\newcommand{\onbb}        {{$0\nu\beta\beta$}}

\newcommand{\gerda}       {\textsc{Gerda}}
  
\newcommand{\gess}        {{$^{76}$Ge}}

\begin{document}


\title{The first search for bosonic super-WIMPs with masses up to 1\,MeV/c$^2$ with GERDA}

\collaboration{{\textsc{Gerda} collaboration}}
\email{correspondence: gerda-eb@mpi-hd.mpg.de}
\noaffiliation

  \affiliation{INFN Laboratori Nazionali del Gran Sasso and Gran Sasso Science Institute, Assergi, Italy}
  \affiliation{INFN Laboratori Nazionali del Gran Sasso and Universit{\`a} degli Studi dell'Aquila, L'Aquila,  Italy}
  \affiliation{INFN Laboratori Nazionali del Sud, Catania, Italy}
  \affiliation{Institute of Physics, Jagiellonian University, Cracow, Poland}
  \affiliation{Institut f{\"u}r Kern- und Teilchenphysik, Technische Universit{\"a}t Dresden, Dresden, Germany}
  \affiliation{Joint Institute for Nuclear Research, Dubna, Russia}
  \affiliation{European Commission, JRC-Geel, Geel, Belgium}
  \affiliation{Max-Planck-Institut f{\"u}r Kernphysik, Heidelberg, Germany}
  \affiliation{Dipartimento di Fisica, Universit{\`a} Milano Bicocca, Milan, Italy}
  \affiliation{INFN Milano Bicocca, Milan, Italy}
  \affiliation{Dipartimento di Fisica, Universit{\`a} degli Studi di Milano and INFN Milano, Milan, Italy}
  \affiliation{Institute for Nuclear Research of the Russian Academy of Sciences, Moscow, Russia}
  \affiliation{Institute for Theoretical and Experimental Physics, NRC ``Kurchatov Institute'', Moscow, Russia}
  \affiliation{National Research Centre ``Kurchatov Institute'', Moscow, Russia}
  \affiliation{Max-Planck-Institut f{\"ur} Physik, Munich, Germany}
  \affiliation{Physik Department, Technische  Universit{\"a}t M{\"u}nchen, Germany}
  \affiliation{Dipartimento di Fisica e Astronomia, Universit{\`a} degli Studi di 
Padova, Padua, Italy}
  \affiliation{INFN  Padova, Padua, Italy}
  \affiliation{Physikalisches Institut, Eberhard Karls Universit{\"a}t T{\"u}bingen, T{\"u}bingen, Germany}
  \affiliation{Physik-Institut, Universit{\"a}t Z{\"u}rich, Z{u}rich, Switzerland}
%
%
\author{M.~Agostini}
  \affiliation{Physik Department, Technische  Universit{\"a}t M{\"u}nchen, Germany}
\author{A.M.~Bakalyarov}
  \affiliation{National Research Centre ``Kurchatov Institute'', Moscow, Russia}
\author{M.~Balata}
  \affiliation{INFN Laboratori Nazionali del Gran Sasso and Gran Sasso Science Institute, Assergi, Italy}
\author{I.~Barabanov}
  \affiliation{Institute for Nuclear Research of the Russian Academy of Sciences, Moscow, Russia}
\author{L.~Baudis}
  \affiliation{Physik-Institut, Universit{\"a}t Z{\"u}rich, Z{u}rich, Switzerland}
\author{C.~Bauer}
  \affiliation{Max-Planck-Institut f{\"u}r Kernphysik, Heidelberg, Germany}
\author{E.~Bellotti}
  \affiliation{Dipartimento di Fisica, Universit{\`a} Milano Bicocca, Milan, Italy}
  \affiliation{INFN Milano Bicocca, Milan, Italy}
\author{S.~Belogurov}
  \altaffiliation[also at: ]{NRNU MEPhI, Moscow, Russia}
  \affiliation{Institute for Theoretical and Experimental Physics, NRC ``Kurchatov Institute'', Moscow, Russia}
  \affiliation{Institute for Nuclear Research of the Russian Academy of Sciences, Moscow, Russia}
\author{A.~Bettini}
  \affiliation{Dipartimento di Fisica e Astronomia, Universit{\`a} degli Studi di 
Padova, Padua, Italy}
  \affiliation{INFN  Padova, Padua, Italy}
\author{L.~Bezrukov}
  \affiliation{Institute for Nuclear Research of the Russian Academy of Sciences, Moscow, Russia}
% ...
%
%
%


\begin{abstract}
We present the first search for bosonic super-WIMPs as keV-scale dark matter candidates performed with the \gerda\ experiment. \gerda\ is a neutrinoless double-beta decay experiment which operates high-purity germanium detectors enriched in $^{76}$Ge in an ultra-low background environment at the Laboratori Nazionali del Gran Sasso (LNGS) of INFN in Italy. Searches were performed for pseudoscalar and vector particles in the mass region from 60\,keV/c$^2$ to 1\,MeV/c$^2$. No evidence for a dark matter signal was observed, and the most stringent constraints on the couplings of super-WIMPs with masses above 120\,keV/c$^2$ have been set. As an example, at a mass of 150\,keV/c$^2$ the most stringent direct limits on the dimensionless couplings of axion-like particles and dark photons to electrons of $g_{ae} <  \text{3}\cdot 10^{-12}$ and ${\alpha'}/{\alpha} < \text{6.5} \cdot 10^{-24}$ at 90\% credible interval, respectively, were obtained. 
\end{abstract}

\maketitle

The evidence for the existence of nonbaryonic dark matter (DM) in our Universe is overwhelming. In particular, recent measurements of temperature fluctuations in the cosmic microwave background radiation yield a 26.4\% contribution of DM to the overall energy density in the $\Lambda$CDM model~\cite{Aghanim:2018eyx}. However, all evidence is gravitational in nature, and the composition of this invisible form of matter is not known. Theoretical models for particle DM yield candidates with a wide range of masses and scattering cross sections with Standard Model (SM) particles~\citep{Bertone:2004pz,Profumo:2019ujg,Tanabashi:2018oca}. Among these, so-called bosonic super-weakly interacting massive particles (super-WIMPs) with masses at the keV-scale and ultra-weak couplings 
to the SM can be cosmologically viable and produce the required relic abundance~\cite{Pospelov:2008jk,An:2014twa}.

Direct DM detection experiments, as well as experiments built to observe neutrinoless double-beta (\onbb) decay   can search for pseudoscalar (also known as axion-like particles, or ALPs) and vector (also known as dark photons) super-WIMPs via their absorption in detector materials in processes analogous to the photoelectric effect (also known as axioelectric effect in the case of axions). The ALP and dark photon energy is transferred to an electron, which deposits its energy in the detector. The expected signature is a full absorption peak in the energy spectrum, corresponding to the mass of the particle, given that these DM candidates have very small kinetic energies~\footnote{Most of the dark matter is cold and non-relativistic, hence $E\simeq m_{DM}$}. For ALPs the coupling to electrons is parameterised via the dimensionless coupling constant $g_{ae}$ \cite{Derevianko:2010kz}, while for dark photons a kinetic mixing $\alpha'$  with strength $\kappa$~\cite{An:2013yua} is introduced in analogy to the electromagnetic fine structure constant $\alpha$, such that ${\alpha'} = (e \kappa)^2/4\pi$. 

The most stringent direct constraints on these couplings for particle masses at the keV-scale are set by experiments using liquid xenon ({\sc{xmass}}, Panda{\sc{x-II}}, {\sc{lux}}, {\sc{xenon100}}, \cite{Abe:2018owy,Fu:2017lfc,Akerib:2017uem,Aprile:2017lqx}), germanium crystals ({\sc{Majorana Demonstrator}}, Super{\sc{cdms}}, {\sc{Edelweiss}}, {\sc{cdex}}, \cite{Abgrall:2016tnn,Aralis:2019nfa,Armengaud:2018cuy,Liu:2016osd}) and calcium tungstate crystals ({\sc{cresst-II}}~\cite{Angloher:2016rji}). Together, these experiments probe the super-WIMP mass region up to $\sim$500\,keV/c$^2$.

Here we describe a search for super-WIMP absorption in the germanium detectors operated by the \gerda\ collaboration, extending for the first time the mass region to 1\,MeV/c$^2$.  At masses larger than twice the electron mass vector particles can decay into $(e^+, e^-)$-pairs and their lifetime would be too short to account for the DM~\cite{An:2014twa}.

The primary goal of \gerda\  is to search for the \onbb\ decay of \gess, deploying High-Purity Germanium (HPGe) detectors enriched up to 87\% in \gess. 
The experiment is located underground at the Laboratori Nazionali del Gran Sasso (LNGS) of INFN, Italy, at a depth of about 3500 meter water equivalent. The HPGe detector array is made of {{7 enriched coaxial and 30 Broad Energy Germanium (BEGe) diodes, with average masses of 2.2\,kg and 667\,g, respectively, leading to a higher full absorption efficiency for the larger coaxial detectors}}. It is operated inside a 64\,m$^3$ liquid argon (LAr) cryostat, which provides cooling and a high-purity, active shield against background radiation. The cryostat is inside a water tank instrumented with PMTs to
 detect Cherenkov light from muons passing through, and thus reduces the muon-induced background to negligible levels. A detailed description of the experiment can be found in~\cite{Agostini:2017hit}, while the most recent \onbb\ decay results are presented in~\cite{Agostini:2019hzm}. 

\begin{figure*}[th!]
\includegraphics[scale=0.45]{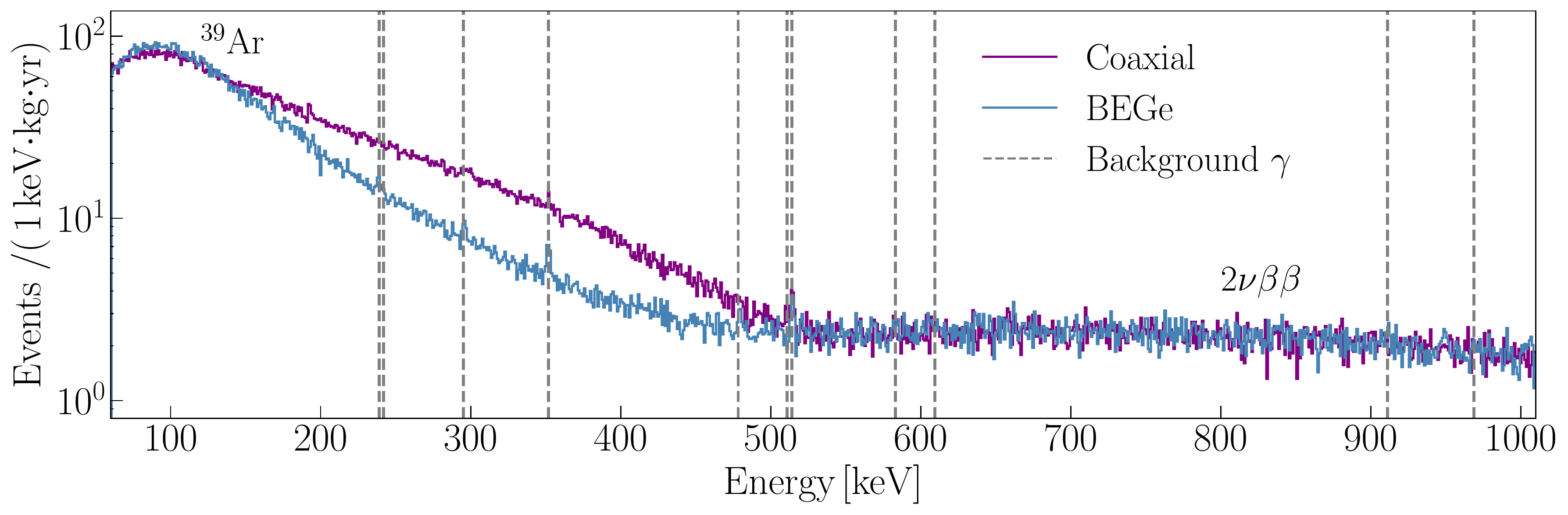}
\caption{The energy spectra of the BEGe and coaxial data sets, normalized by exposure. Only events with energies up to 1\,MeV were considered in the analysis. The coaxial data set shows a significantly higher event rate (mainly from $^{39}$Ar decays) at energies below 500\,keV due to the larger surface area of the signal readout electrodes~\cite{Agostini:2019hzm}.
The dashed lines indicate the positions of the main known background gamma lines, also listed in Table~\ref{tab:gammas}. }
\label{fig:zom_lowen}
\end{figure*}

Due to its ultra-low background level~\cite{Abramov:2019hhh} and excellent energy resolution ($\sim$3.6\,keV and $\sim$3.0\,keV full width at half maximum (FWHM) for coaxial and BEGe detectors at \Qbb=2039\,keV, respectively), the \gerda\ experiment is well-suited to search for other rare interactions, in particular for peak-like signatures as expected from bosonic super-WIMPs. Here we make the assumption that super-WIMPs constitute all of the DM in our galaxy, with a local density of 0.3\,GeV/cm$^3$~\cite{Green:2017odb}. The absorption rate for dark photons and ALPs in an Earth-bound detector can be expressed  as \cite{Pospelov:2008jk}:

\begin{equation}
\label{eq:vec_rate}
 R \approx \frac{4\cdot 10^{23}}{A}\frac{\alpha'}{\alpha}\left(\frac{\mathrm{[keV/c^2]}}{m_v}\right)\left(\frac{\sigma_{pe}}{\mathrm{[b]}}\right)\mathrm{kg^{-1}d^{-1}}
\end{equation} and
\begin{equation}
\label{eq:scal_rate}
R \approx\frac{1.2\cdot 10^{19}}{A}g_{ae}^{2}\left(\frac{m_a}{\mathrm{[keV/c^2]}}\right)\left(\frac{\sigma_{pe}}{\mathrm[{\text{b}}]}\right)\mathrm{kg^{-1}d^{-1}}~,
\end{equation}
respectively.
Here $g_{ae}$ and ${\alpha'}/{\alpha}$ are the dimensionless coupling constants, 
\textit{A} is the atomic mass of the absorber, $\sigma_{pe}$ is the photoelectric cross section 
on the target material (germanium), and $m_v$ and $m_a$ are the DM particle masses. 
{{The linear versus inverse proportionality of the rate with the particle mass is due to the 
fact that rates scale as flux times cross section, where the cross section is proportional 
to $m_a^2$ and ${\alpha'}/{\alpha}$ in the pseudoscalar and 
vector boson case, respectively~\cite{Pospelov:2008jk}.}}

We perform the search for super-WIMPs in the (200\,keV/c$^2$ - 1\,MeV/c$^2$) mass range on data collected between December\,2015 and April\,2018,  corresponding to 58.9\,\kgyr\  of exposure. The energy threshold of the HPGe detectors was lowered in October 2017 and enabled a search in the additional mass range of (60\,keV/c$^2$ - 200\,keV/c$^2$), corresponding to 14.6\,\kgyr\ of exposure accumulated until April\,2018. The individual exposures for BEGe and coaxial detectors above (below) 200\,keV are 30.8~(7.7) and 28.1~(6.9)\,\kgyr, respectively. The lower energy bounds for our analysis were motivated by {the energy thresholds of the Ge detectors} and the shape of the background spectrum (dominated by $^{39}$Ar decays) and the size of the fit window as explained in the following.

In \gerda\ the energy reconstruction of events is performed through digital pulse processing~\cite{Agostini:2015pta}. Events of non-physical origin such as discharges are rejected by a set of selection criteria based on waveform parameters (i.e., baseline, leading edge, and decay tail). The efficiency of these cuts for accepting signal events was estimated at $>$~98.7\%. Since super-WIMPs would interact only once in a HPGe diode, events tagged in coincidence with the muon or LAr vetos, or observed in more than one germanium detector, were rejected as due to background interactions. We use the same set of cuts as in the \gerda\ main analysis for the \onbb\ decay~\cite{Agostini:2019hzm}, with the exception of the pulse shape discrimination cut, which had been tailored to the high-energy \onbb\ decay search. The {{muon and}} LAr veto {{accept}} signal events with {{efficiencies  of 99.9\%~\cite{Freund:2016fhz} and 97.5\%~\cite{Agostini:2019hzm}, respectively}}. 

The total efficiency to observe a super-WIMP absorption in the HPGe diodes was  determined as:

\begin{equation}
\epsilon_\text{tot} = \epsilon_\text{cuts}\frac{1}{\mathcal{E}} \sum_i^{N_\text{det}}\mathcal{E}_\text{i} \cdot f_\text{av,i}\cdot\epsilon_\text{fep,i}~,
\label{eq:eq_corr_rate}
\end{equation}
where the efficiency of the event selection criteria $\epsilon_\text{cuts}$ and the exposure $\mathcal{E}$  of each data set  were taken into account. The index \textit{i} runs over the individual detectors of that data set, containing $N_\text{det}$ detectors, $\mathcal{E}_i$ is the exposure, $f_\text{av,i}$ the active mass fraction, and $\epsilon_\text{fep,i}$ the  efficiency for detection of the full energy absorption of an electron emitted in the interaction. With the exception of $\epsilon_\text{fep,i}$ all parameters were identical to those in the analysis presented in\,\cite{Agostini:2019hzm}. The full-energy absorption efficiency  $\epsilon_\text{fep,i}$ accounts for partial energy losses, for example  in a detector's dead layer. This efficiency was estimated for each detector at energies between 60 and 1000\,keV with a Monte Carlo simulation of uniformly distributed electrons in the active volume of the detector using the MaGe framework \cite{Boswell:2010mr}. 

Table~\ref{tab:effsummary} shows the average full energy absorption detection efficiencies $\epsilon_\text{fep}$ and the total efficiencies $\epsilon_\text{tot}$ at 
the lower and upper boundaries of the search region. At 60\,keV, the full energy absorption  was estimated as 99.5\% for all detectors, while at 1000\,keV 
it is  95.1\% and 96.2\% on average for BEGe and coaxial detectors, respectively. {{The energy dependence of the efficiency is caused by the 
photoabsorption cross section and the different size of the germanium diodes}}.
The events which survived all selection criteria (with total efficiencies between 85.7\% and 81.4\%, see Table~\ref{tab:effsummary}) are shown in Fig.~\ref{fig:zom_lowen} for the coaxial and BEGe detector data sets.

\begin{table}[h!]
\center
\caption{Detection efficiencies for the super-WIMP search. The average $ \langle \epsilon_\text{fep} \rangle$ value for the detectors from one data set and the total efficiency $\epsilon_\text{tot}$ are shown at  60\,keV and 1000\,keV for the two data sets.}
\begin{tabular}{c|cc}
& BEGe & Coaxial \\ \hline
$ \langle \epsilon_\text{fep} \rangle\rvert$E=\,60\,keV & 99.5\% & 99.5\% \\
$ \langle \epsilon_\text{fep} \rangle\rvert$E=\,1000\,keV & 95.1\% & 96.2 \% \\
$\epsilon_\text{tot}\rvert$E=\,60\,keV &   85.7\%  & 84.2\% \\ 
$\epsilon_\text{tot}\rvert$E=\,1000\,keV & 82.0\% & 81.4\%  \\
\end{tabular}
\label{tab:effsummary}
\end{table}

\begin{table}[h!]
\caption{Gamma lines accounted for in the background model for the super-WIMP search (branching ratio above 0.1\%).}
\vspace{0.3cm}
\begin{tabular}{c| l}
Isotope of origin & Energy\,[keV]\\ \hline
{\rule{0mm}{5mm}}
$^{228}$Ac &  478.3, 911.2,  969.0 \\
$^{208}$Tl &  583.2\\
$^{212}$Pb & 238.6 \\
$^{214}$Pb & 242.0,  295.2, 352.0 \\
$^{214}$Bi & 609.3\\
$^{85}$Kr &  514.0\\
e$^{+}e^{-}$-annih. & 511.0\\
\end{tabular}
\label{tab:gammas}
\end{table}

The expected signal from super-WIMPs has been modeled with a  Gaussian peak broadened by the energy resolution of the HPGe detectors. 
To estimate the potential signals from these particles we performed a binned Bayesian fit (with a 1\,keV binning, while the systematic uncertainties on the energy scale are estimated at 0.2\,keV) of the signal and a background model of the data. The fit was performed within a window of 24\,keV in width, centered on the energy corresponding to the hypothetical mass of the particle and sliding with 1\,keV step to examine each mass value. The total number of counts from signal and background was determined as follows:
\begin{equation}
\label{eq:model}
R_\text{tot}(E) =  G_0(\mathcal{N}_0, E_0, \sigma_0) +  F(E) + G_\gamma( \mathcal{N}_\gamma, E_\gamma, \sigma_\gamma )~,
\end{equation}
where the Gaussian function $G_0$ models the peak signal of super-WIMPs at a fixed energy $E_0$, corresponding to their mass. 
The Gaussian $G_\gamma$ models the background gamma lines with energy  $E_{\gamma}$ listed in Table \ref{tab:gammas} in case it is found within the sliding fit window. For more than one background gamma line the Eq.~\ref{eq:model} is modified accordingly to model all the peaks. $\mathcal{N}_0$ and $\mathcal{N}_{\gamma}$ are the counts in the fitted signal and background peaks, respectively. The effective energy resolutions $\sigma_0$ and $\sigma_\gamma$ of the detectors from the combined spectra are fixed to the values obtained from the regularly acquired calibration data, with systematic uncertainties around 0.1\,keV~\cite{Agostini:2019hzm}. Finally, the polynomial fit function $F(E)$ describes the continuous background, and was chosen as a first- and second-order polynomial for energies above and below 120\,keV, respectively. The higher order polynomial at lower energies is motivated by the curvature  of the $^{39}$Ar beta spectrum, see Fig.~\ref{fig:zom_lowen}. At other energies, the spectrum has an approximately linear shape, and thus a first-order polynomial was judged sufficient.

The Bayesian fit was performed with the BAT framework \cite{BAT} using the Markov Chain Monte Carlo  technique \cite{MCMC} to compute the marginalized posterior probability density function (PDF) given energy values of the data $\mathbf{E}$, $P(R_\text{S}| \mathbf{E})$, where $R_S$ is the signal rate, i.e., the number of counts normalized by exposure.

The probability for the signal count rate $P\left(R_S, \boldsymbol{\theta} | \mathbf{E}, M\right)$, 
given data $\mathbf{E}$ and a model M, is described by Bayes' theorem as:
\begin{equation}
 P\left(R_S, \boldsymbol{\theta} | \mathbf{E}, M\right)=
\\  \frac{ P\left(\mathbf{E} | R_S, \boldsymbol{\theta}, M\right) 
\pi\left(R_S\right) \pi(\boldsymbol{\theta})}
{ \int \int P\left(\mathbf{E} | R_S, \boldsymbol{\theta}, M\right) 
\pi\left(R_S\right) \pi(\boldsymbol{\theta}) d \boldsymbol{\theta} d R_S }.
\end{equation}
The denominator defines the overall probability of obtaining the observed data given a hypothetical signal. The numerator includes prior probabilities $\pi$ for the signal count rate $R_S$ and for the nuisance parameters $\boldsymbol{\theta}$ (e.g., background shape) estimated before performing the fit. For $\boldsymbol{\theta}$, flat priors were adopted, bound generously according to a preliminary fit with the Minuit algorithm~\cite{minuit}. For the signal count rate, $R_S$, the uniform (i.e., constant over the defined range) prior probability was constructed to be positive, with the upper bound defined by the total number of events in the signal region plus 10 times the expected Poisson fluctuations. The conditional probability $P\left(\mathbf{E} | R_S, \boldsymbol{\theta}, M\right)$ is estimated according to the super-WIMP interaction model given by Eq.~\ref{eq:model} and Poisson fluctuations in the data. 

\begin{figure}
\center
\includegraphics[width=0.5\textwidth]{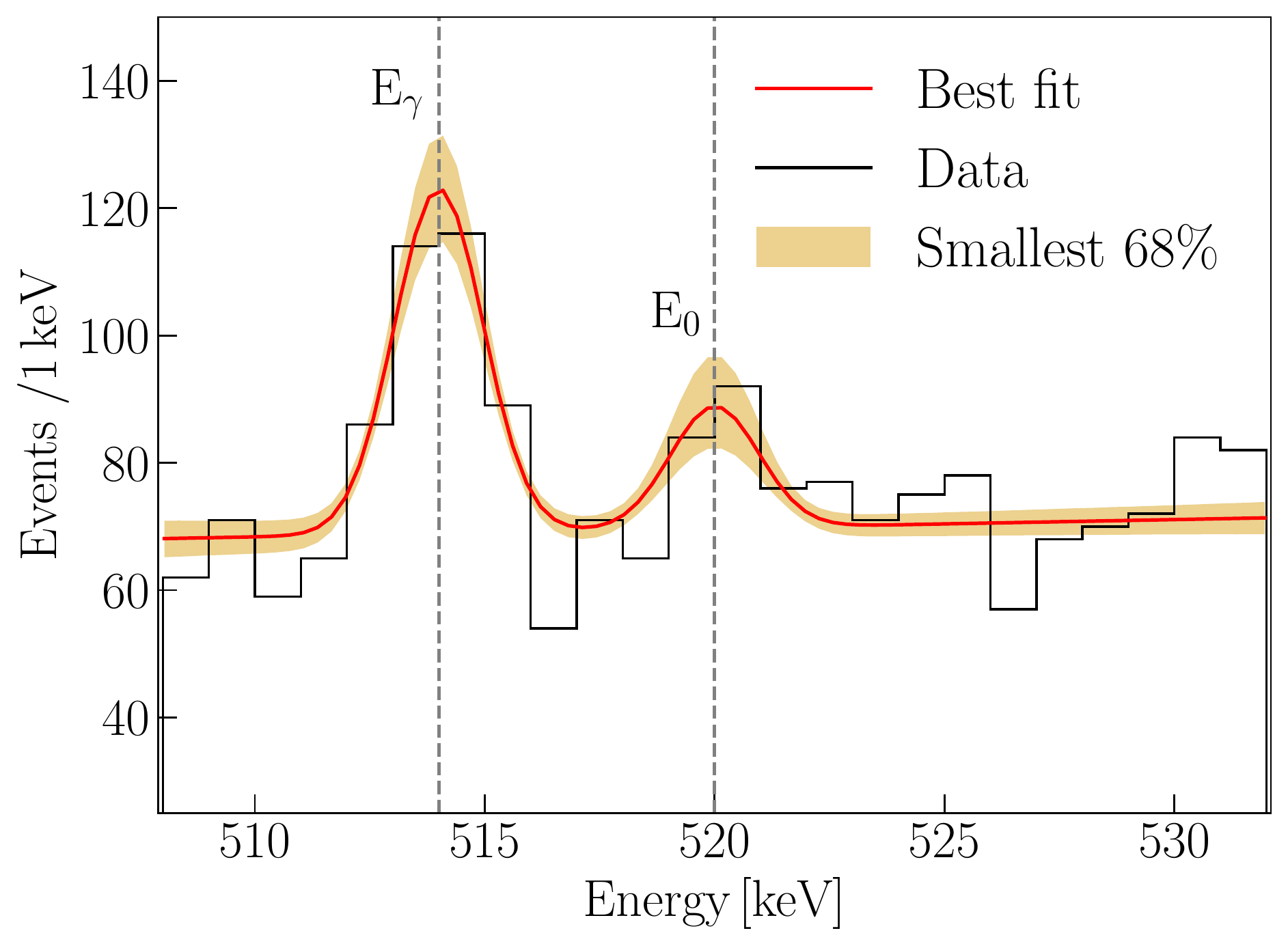} 
\includegraphics[width=0.5\textwidth]{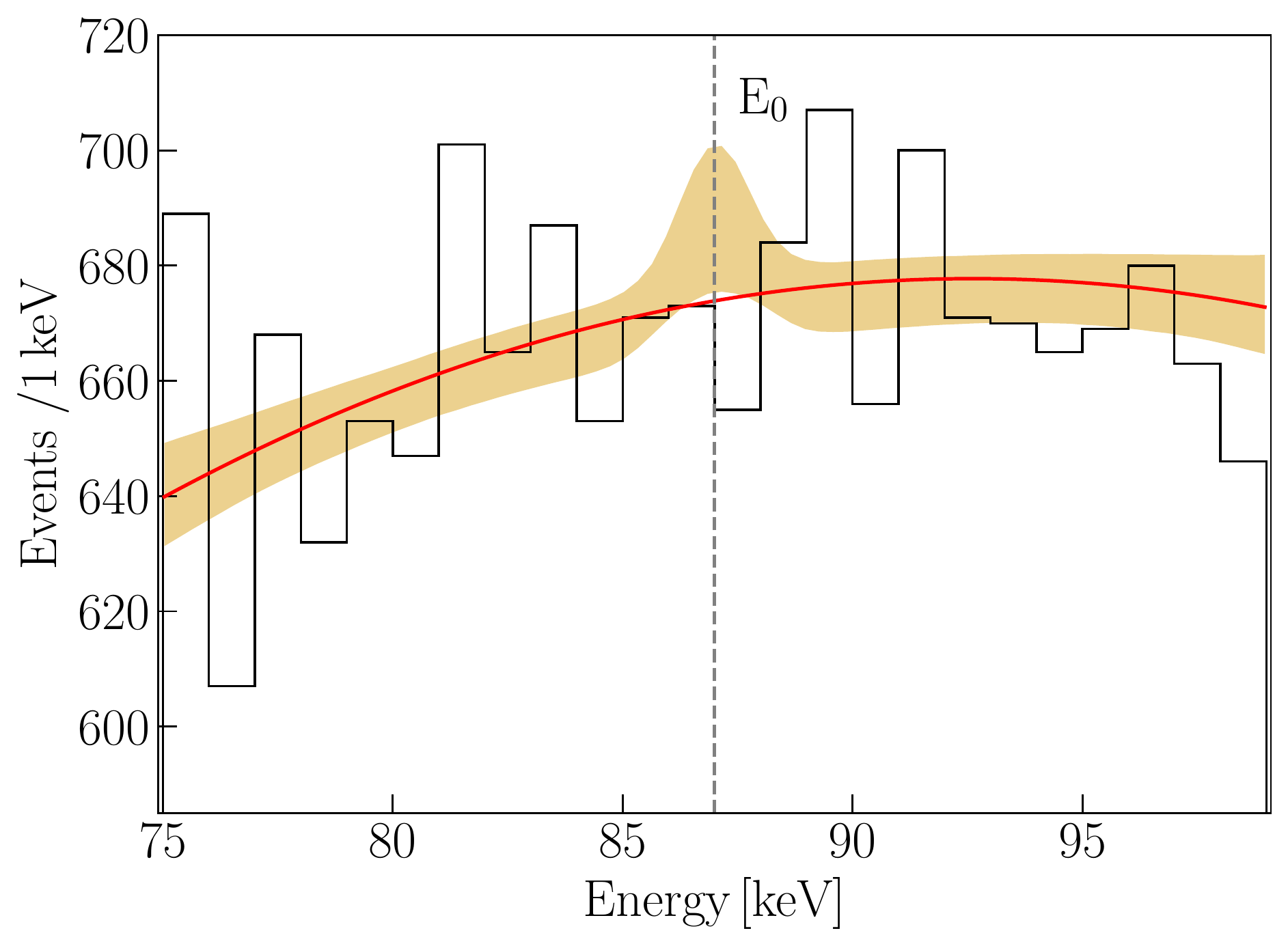} 
\caption{ Best fit (red lines) and 68\% uncertainty band (yellow bands) from marginalised posterior PDFs of the model parameters assuming a hypothetical signal at E$_0$ in the BEGe data set. \textit{Top}: Fit of a signal assumed at E$_0$=520\,keV; the excess is at a level of 2.6\,$\sigma$. A first order polynomial is used for the continuous background and a Gaussian for the background gamma line due to the decay of $^{85}$Kr. \textit{Bottom}: Fit of a signal assumed at E$_0$=87\,keV, using a second order polynomial for the continuous background.
}
\label{fig:fit_res}
\end{figure}

The reported results were obtained from the combined fit of the BEGe and coaxial data sets. {{First, the BEGe data set was fit using a flat prior for the signal count rate R$_S$, and the obtained posterior was used as a prior for the fit of the Coaxial data set. The results of the latter were then employed to evaluate the corresponding coupling constants of the super-WIMPs.}} The detection of the signal is ruled out when the significance of the best fit value for the count rate is less than 5 sigma, estimated as half of the 68\% quantile of the posterior PDF. Additionally, if a fitted signal is in close proximity (within 5~$\sigma$ of the energy resolution) to a known background gamma line, an upper limit was set irrespective of the mode,  as uncertainties in the background rate do not allow to reliably claim an excess signal above gamma lines.
An example for the fit using the model described by Eq.~\ref{eq:model} for two different background functions $F(E)$ is shown in Fig.~\ref{fig:fit_res}.

The obtained posterior PDFs do not show evidence for a signal in the energy range of the analysis. We thus set 90\% credible interval (C.I.) upper limits  on the signal count rate, corresponding to  the 90\% quantile of the posterior PDF $P(R_\text{S}| \mathbf{E})$, accounting for the detection efficiencies according to Eq.~\ref{eq:eq_corr_rate}.

\begin{figure}
\centering
\includegraphics[width=0.5\textwidth]{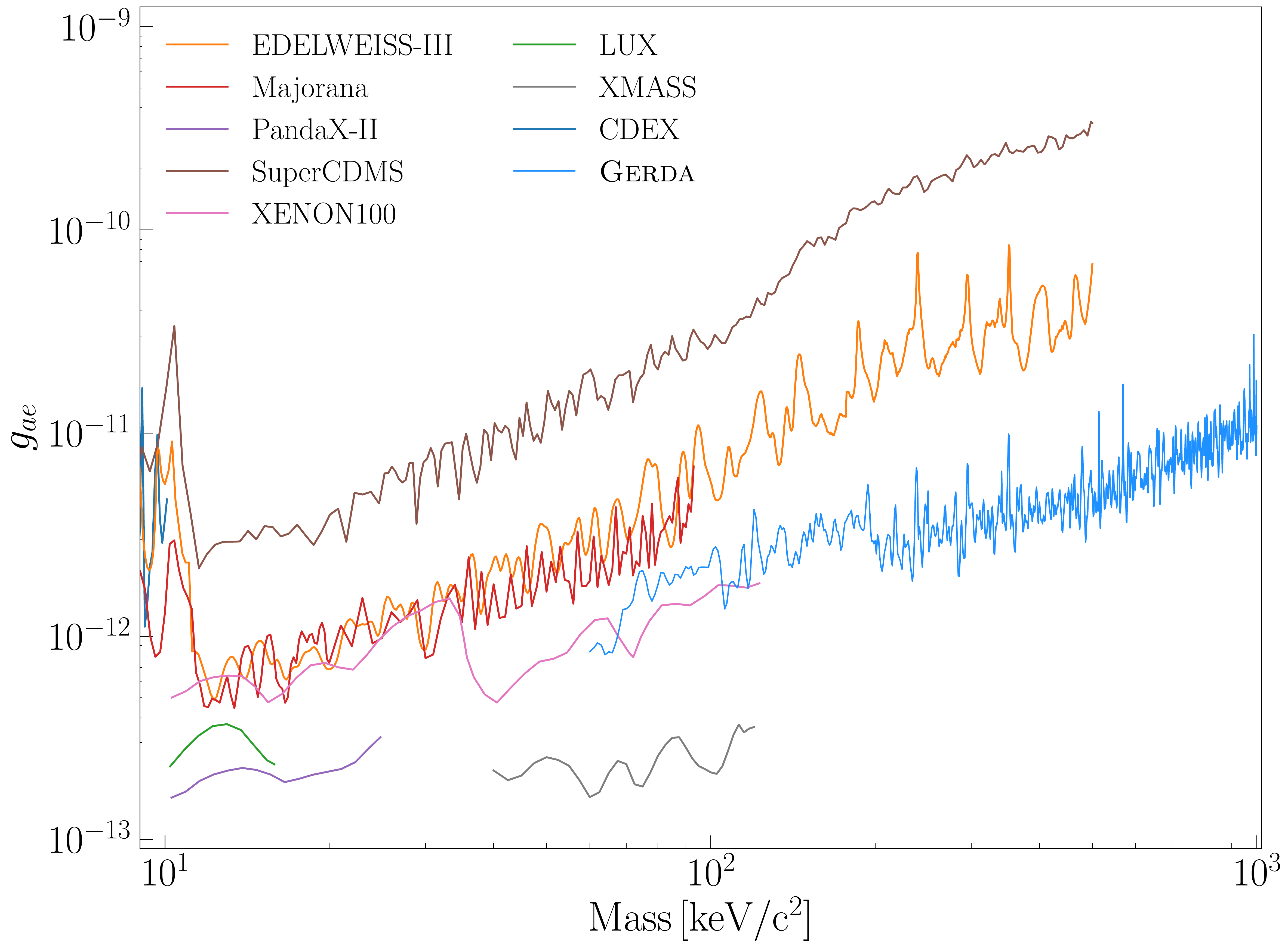}
\includegraphics[width=0.5\textwidth]{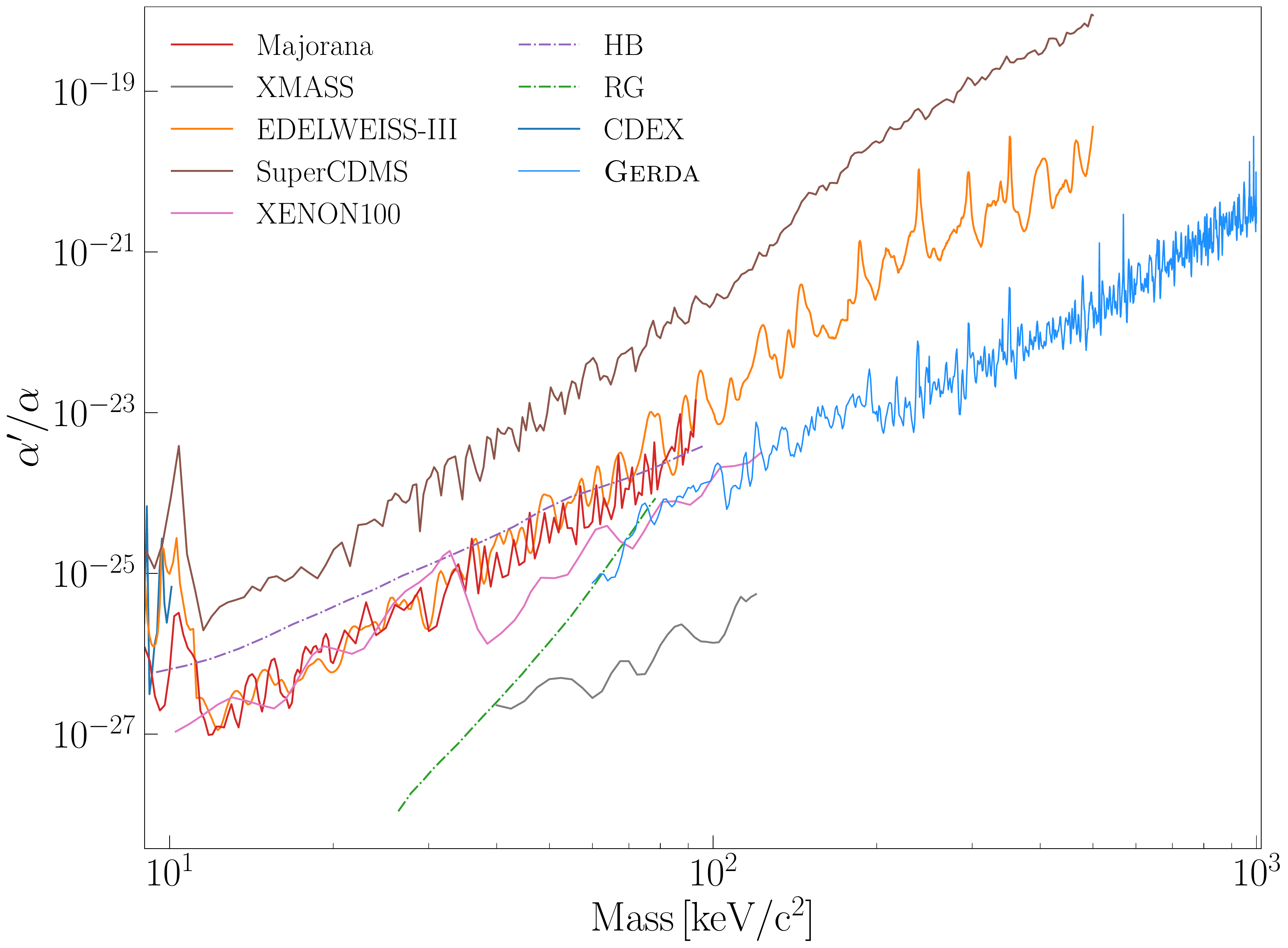}
\caption{Upper limits (at 90\% C.I.)  on the coupling strengths of pseudo-scalar (top) and vector (bottom) super-WIMPs. 
Only part of the data was acquired with a lower energy threshold, resulting in a lower exposure for data below 200\,keV/c$^2$ and causing the step-like feature around this energy. Results from other experiments (see text) are also shown, together  with indirect constraints from anomalous energy losses in horizontal branch (HB) and red giant (RG) stars (we refer to~\cite{An:2014twa} for details). }
\label{fig:coupl_consts}
\end{figure}

The 90\% C.I. limits on the signal rate $R_S$ were converted into upper limits on the coupling strengths using Eqs.~\ref{eq:vec_rate} and \ref{eq:scal_rate}. The results are presented in Fig.~\ref{fig:coupl_consts}.  We compare these to  direct detection limits from {\sc{cdex}}~\cite{Liu:2016osd}, {\sc{Edelweiss}}-III~\cite{Armengaud:2018cuy}, {\sc{lux}}~\cite{Akerib:2017uem}, the {\sc{Majorana Demonstrator}}~\cite{Abgrall:2016tnn}, Panda{\sc{x}}-II~\cite{Fu:2017lfc}, Super{\sc{cdms}}~\cite{Aralis:2019nfa}, {\sc{xenon100}}~\cite{Aprile:2017lqx} and  {\sc{xmass}}~\cite{Abe:2018owy}, as well as to indirect limits from horizontal branch and red giant stars \cite{Pospelov:2008jk}. Above 120\,keV/c$^2$ indirect $\alpha'/\alpha$ limits from decays of vector-like particles into three photons ($V\rightarrow 3\gamma$) are
 significantly lower (ranging from 10$^{-12}$ at masses of 100\,keV/c$^2$ to 10$^{-16}$ at 700\,keV/c$^2$) than the available direct limits (not shown)~\cite{An:2014twa}. {{The improvement in sensitivity with respect to other crystal-based experiments is due to the much larger exposure in \gerda\ and the lower background rate over all of the search region.}}

The weakening of our upper limits with increasing mass is primarily
due to the steep decrease of the photoelectric cross section from about 45\,barn at 100\,keV to 0.085\,barn at 1\,MeV that overrules both the linear and inverse mass dependence in Eqs.~\ref{eq:vec_rate} and \ref{eq:scal_rate}. {{The fluctuations in the upper limit curves are due to background fluctuations, where prominent peaks come from known gamma lines, shown in Table~\ref{tab:gammas}}}.

To summarise, in this Letter we demonstrated the capability of \gerda\ to search for other rare events besides the \onbb\ decay of $^{76}$Ge. We  performed a search for keV-scale DM  in the form of bosonic super-WIMPs based on data with exposures of 58.9\,\kgy\ and 14.6\,\kgy\ in the mass ranges of (200\,keV/c$^2$-1\,MeV/c$^2$) and (60\,keV/c$^2$-200\,keV/c$^2$), respectively. Upper limits on the coupling strengths $g_{ae}$ and $\alpha'/\alpha$ were obtained from a Bayesian fit of a background model and a potential peak-like signal to the measured data. Our limit is compatible with other direct searches in the mass range (60\,keV/c$^2$-120\,keV/c$^2$) where the strongest limits were obtained by xenon-based DM experiments due to higher exposures and lower background rates in this low-energy region. Our search probes for the first time the mass region up to 1\,MeV/c$^2$ and sets the best direct constraints on the couplings of super-WIMPs over a large mass range from (120\,keV/c$^2$-1\,MeV/c$^2$). As an example, at a mass of 150\,keV/c$^2$ the most stringent direct limits on the dimensionless couplings of axion-like particles and dark photons to electrons of $g_{ae} <  \text{3}\cdot 10^{-12}$ and ${\alpha'}/{\alpha} < \text{6.5} \cdot 10^{-24}$ (at 90\% C.I.), respectively, were established. The limits are affected by the known background gamma lines, listed in Table~\ref{tab:gammas}, due to higher background rate at these energies. 

The sensitivity to new physics is expected to improve in the near future with the upcoming {\sc{Legend}}-200 experiment. The experimental program aims to decrease the background rate and increase the number of HPGe detectors operated in an upgraded \gerda\ infrastructure at LNGS~\cite{Abgrall:2017syy}.

The \gerda\ experiment is supported financially by the Swiss National Science Foundation
(SNF), German Federal Ministry for Education and Research
(BMBF), the German Research Foundation (DFG) via the Excellence Cluster Universe and the SFB1258, the
Italian Istituto Nazionale di Fisica Nucleare (INFN), the
Max Planck Society (MPG), the Polish National Science
Centre (NCN), the Foundation for Polish Science (TEAM/
2016-2/2017), and the Russian Foundation for Basic Research
(RFBR). The institutions acknowledge also internal financial
support. This project has received funding or support from
the European Union’s Horizon 2020 research and innovation programme under the Marie Sklodowska-Curie Grant
Agreements No. 690575 and No. 674896, respectively. The
\gerda\ collaboration thanks the directors and the staff
of the LNGS for their continuous strong support of the
\gerda\ experiment.

\bibliographystyle{apsrev4-1}

\addcontentsline{toc}{chapter}{Bibliography}


%




\end{document}